\documentclass[11pt,a4paper]{article}
\usepackage{amsmath,slashed,amssymb,epsfig,amsthm,bm,graphicx,hyperref,tensor}
\usepackage[caption=false]{subfig}
\usepackage{color}
\usepackage[top=2cm, bottom=2cm, left=3.0cm, right=3.0cm]{geometry}
\usepackage[affil-it]{authblk}

\usepackage[backend=bibtex,natbib=true,style=phys,articletitle=false,biblabel=brackets,giveninits=true,abbreviate=true,doi=false,url=false,isbn=false,eprint=true]{biblatex}
\definecolor{darkgreen}{rgb}{0,0.35,0}

\newcommand{\ii}{\ensuremath{\mathrm{i}}}
\newcommand{\christoffel}[3]{\left\{
\begin{array}{c}
#1\\ #2 \, #3
\end{array}
\right\}}

\newcommand{\UACh}{Instituto de Ciencias F\'isicas y Matem\'aticas, Universidad Austral  de Chile, Casilla 567, 5090000 Valdivia, Chile}
\newcommand{\UCharles}{IPNP - Faculty of Mathematics and Physics, Charles University, V Hole\v{s}ovi\v{c}k\'ach 2, 18000 Prague 8, Czech Republic}

\begin{document}

\title{Time-loops to spot torsion on bidimensional Dirac materials with dislocations}

\author[1]{Alfredo Iorio\thanks{iorio@ipnp.troja.mff.cuni.cz}}
\author[2,1]{Pablo Pais\thanks{pais@ipnp.troja.mff.cuni.cz}}

\affil[1]{\UCharles}
\affil[2]{\UACh}

\date{}

\maketitle

\begin{abstract}
Assuming that, with some care, dislocations could be meaningfully described by torsion, we propose here a scenario based on a previously unexplored role of time in the low-energy Dirac field theory description of two-space-dimensional Dirac materials. Our approach is based on the realization of an exotic time-loop that could be realized by oscillating particle-hole pairs, overcoming the well-known geometrical obstructions due to the lack of a third spatial dimension. General symmetry considerations allow concluding that the effects we are looking for can only be seen if we move to the nonlinear response regime.
\end{abstract}


\section{Introduction}\label{SecIntroduction}

Einstein’s theory of General Relativity (GR) is the accepted theory of gravitation, surely for its beauty, but perhaps more solidly for its many observational successes. Recent remarkable examples of those are the discovery by the Event Horizon Telescope (EHT) of astrophysical black holes, first in the centre of the galaxy M87 \cite{Event_Horizon_Telescope_Collaboration_2019}, then in the centre of our Milky Way \cite{Event_Horizon_Telescope_Collaboration_2022}, and the observation of the gravitational waves by the Laser Interferometer Gravitational-wave Observatory (LIGO) \cite{Abbott1916}, opening the way to a new era in the exploration of the signals coming from the Universe surrounding us.

However, astronomical observations pointing to dark matter and dark energy, on the one hand, and theoretical challenges that mostly have to do with a quantum theory of gravity, on the other hand, call for a theory that extends GR. Since matter is characterized, at the fundamental level, by the \emph{mass} and the \emph{spin} of its elementary constituents\footnote{As fundamental particles locally obey quantum mechanics and special relativity theory, they are classified by reducible unitary representations of the Poincaré group \cite{Hehl1976}, labelled by mass $m$ and spin $s$.}, and since GR is a geometric theory entirely based on curvature, whose source is the mass, a natural extension of GR is obtained by including torsion in the geometrical properties of the spacetime. Indeed, spin is related to torsion, as mass is related to curvature. On this, the literature is vast, so let us point to the seminal work of Tom Kibble \cite{Kibble1961}.

Therefore, one direction to explore for extensions of GR is to include torsion as an extra feature of the spacetime geometry. This is the case of the Einstein-Cartan (EC) theory of gravitation, where through a contorsion in the spin-connection, torsion appears in the physical phenomena, allowing the possibility of a direct extension of GR. On this, see the two excellent reviews \cite{Hehl1976} and \cite{Shapiro}, which have complementary points of view.

At this time, though, there is no direct experimental evidence of the effects of spacetime torsion, even though some argue that the very existence of spin itself is the only manifestation of spatiotemporal torsion \cite{HehlObukhov2007} and that serious work is currently undertaken within certain cosmological scenarios \cite{NickUniverse2021}. Therefore, it is handy that certain analog systems might furnish tabletop experimental grounds to test fundamental theories of gravity based on torsion.

Indeed, to include torsion in the geometric description of spacetime, Cartan himself was inspired by Cosserat's theory of crystal dislocations, see, e.g., the story told in \cite{HehlObukhov2007}. On the other hand, three-dimensional gravity models have been employed to describe the (un)elastic properties of crystalline materials, where disclinations and dislocations correspond to curvature and torsion in the continuum limit, respectively, see, e.g., \cite{Katanaev2005}.

Given the natural occurrence of topological defects in graphene \cite{pacoreview2009,deJuan2010} and other Dirac materials \cite{WehlingDiracMaterials2014}, and given that such materials have been proposed as versatile analogs of classical and quantum gravitational phenomena, as told in~\cite{i2,iorio2012,iorio2014,iorio2015,ipfirst,CommentPRB2022,ip3,reach,GUPBTZ,threelayers,ip2} and in the recent review \cite{AIPS2022}, then it is very natural to seek for the role of torsion in the analog gravitational physics realized there. However, early attempts in that direction (that we shall recall later in Section \ref{SecRiemannCartanSpaces}) had negative outcomes due to the geometrical obstruction existing in two-space-dimensional materials \cite{deJuan2010}.

Here we present a possible way out from that obstruction, based on the results of \cite{ip4}, where time plays an unexpected role in the $(2+1)$-dimensional Dirac-like description of the low-energy quasiparticles of graphene \cite{pacoreview2009}.

\section{The Riemann-Cartan spaces}\label{SecRiemannCartanSpaces}

Einstein himself \cite{Pantaleo1955cinquant} points out that, for a theory of gravitation as GR, more important than the metric is the parallel transport, taken into account by the affine connection. An $n$-dimensional (differential) manifold $M$, equipped with a connection $\Gamma$, is said to be \emph{linearly connected}, and it is denoted by $L_{n}$.
Given a point $p$ on $M$, we say that a vector $v=v^{\mu}\partial_{\mu}$ of the tangent space at that point, $T_{p}M$, is parallel-transported along the direction $dx^{\nu}$ if its variation is given by
\begin{equation*}
\delta_{\|}v^{\mu} = v^{\mu}_{\|}(x + dx) - v^{\mu}(x) = -\Gamma{^{\mu}}{_{\rho\nu}}\,v^{\nu}\,dx^{\rho} \;,
\end{equation*}
where $\Gamma_{\rho\nu}^{\mu}$ is a \emph{linear connection}. The covariant derivative $\nabla\,v$ is defined by the difference
\begin{equation}
v^{\mu}(x + dx) - v^{\mu}_{\|} (x + dx) =  v^{\mu}(x+dx) - v^{\mu}(x) - \delta_{\|}\,v^{\mu} = \left(\partial_{\nu} v^{\mu} + \Gamma{^{\mu}}{_{\nu\rho}} v^{\rho}\right)\,dx^{\nu} \equiv (\nabla_{\nu} v^{\mu} )\,dx^{\nu}  \;,
\end{equation}
where in the second equality we expanded $v^{\mu}(x + dx)$ just up linear order in $dx$ and, in the third equality, the order of the covariant indices of $\Gamma$ must be noted. We stress here that the connection transforms under coordinate changes in an appropriate way in order for the quantity $\nabla_{\nu} v^{\mu}$ to transform as a covariant vector, and this can be generalized to an arbitrary rank tensor. Henceforth, the difference
\begin{equation}\label{torsion_def}
\tensor{T}{^{\mu}}{_{\nu\rho}} \equiv \left(\Gamma{^{\mu}}{_{\nu\rho}} - \Gamma{^{\mu}}{_{\rho\nu}}\right) \;,
\end{equation}
does transform as a third-rank tensor, called the \emph{torsion tensor}.

Parallel transport is a path-dependent process, and if we parallel transport a vector around an infinitesimal loop, the components change in a way reflected by the \emph{curvature tensor}
\begin{equation}\label{curvature_def}
\tensor{R}{^{\mu}}{_{\sigma\nu\rho}} \equiv \partial_{\nu}\Gamma{^{\mu}}{_{\rho\sigma}} - \partial_{\rho}\Gamma{^{\mu}}{_{\nu\sigma}} + \Gamma{^{\mu}}{_{\nu\tau}}\,\Gamma{^{\tau}}{_{\rho\sigma}} - \Gamma{^{\mu}}{_{\rho\tau}}\,\Gamma{^{\tau}}{_{\nu\sigma}} \;,
\end{equation}
that is a fourth-rank tensor. Importantly, both curvature and torsion can be defined in a manifold without the notion of a metric.

From definitions (\ref{torsion_def}) and (\ref{curvature_def}), it can be shown \cite{Santalo} that curvature and torsion are responsible for the noncommutativity of covariant derivatives of a vector in $L_{n}$,
\begin{equation*}
\nabla_{\nu} \nabla_{\rho} v^{\mu} - \nabla_{\rho} \nabla_{\nu} v^{\mu} = R{^{\mu}}{_{\sigma\nu\rho}}\,v^{\sigma} - \tensor{T}{^{\sigma}_{\nu\rho}} \nabla_{\sigma} v^{\mu} \;,
\end{equation*}
while for a scalar, $\varphi$, the noncommutativity is entirely due to torsion alone
\begin{equation*}
\nabla_{\nu} \nabla_{\rho} \varphi - \nabla_{\rho} \nabla_{\nu} \varphi = - \tensor{T}{^{\sigma}}{_{\nu\rho}} \partial_{\sigma} \varphi \;.
\end{equation*}

It is when we want to make contact with the experiments, and measure angles and distances between events in a spacetime manifold, that we introduce the \emph{ metric tensor} as a second-rank tensor defining the line element, i.e. the infinitesimal distance between two points as
\begin{equation}\label{metric_def}
ds^{2} = g_{\mu\nu}\,dx^{\mu}\,dx^{\nu} \;,
\end{equation}
whereas, by integration, we can define the longitude of any curve on $L_{n}$.

All the above also makes contact with the equivalence principle, as we can measure in inertial spaces \emph{locally} diffeomorphic to Minkowski flat space. This principle was initially formulated by taking into account just point-like test particles and does not consider internal structures as spin. Therefore, in more general settings, the equivalence principle can be formulated in a $U_{n}$ space. On this, see the discussion in \cite{VonDerHeyde1975}.

\begin{figure}
\begin{center}
\includegraphics[width=0.40\textwidth,angle=0]{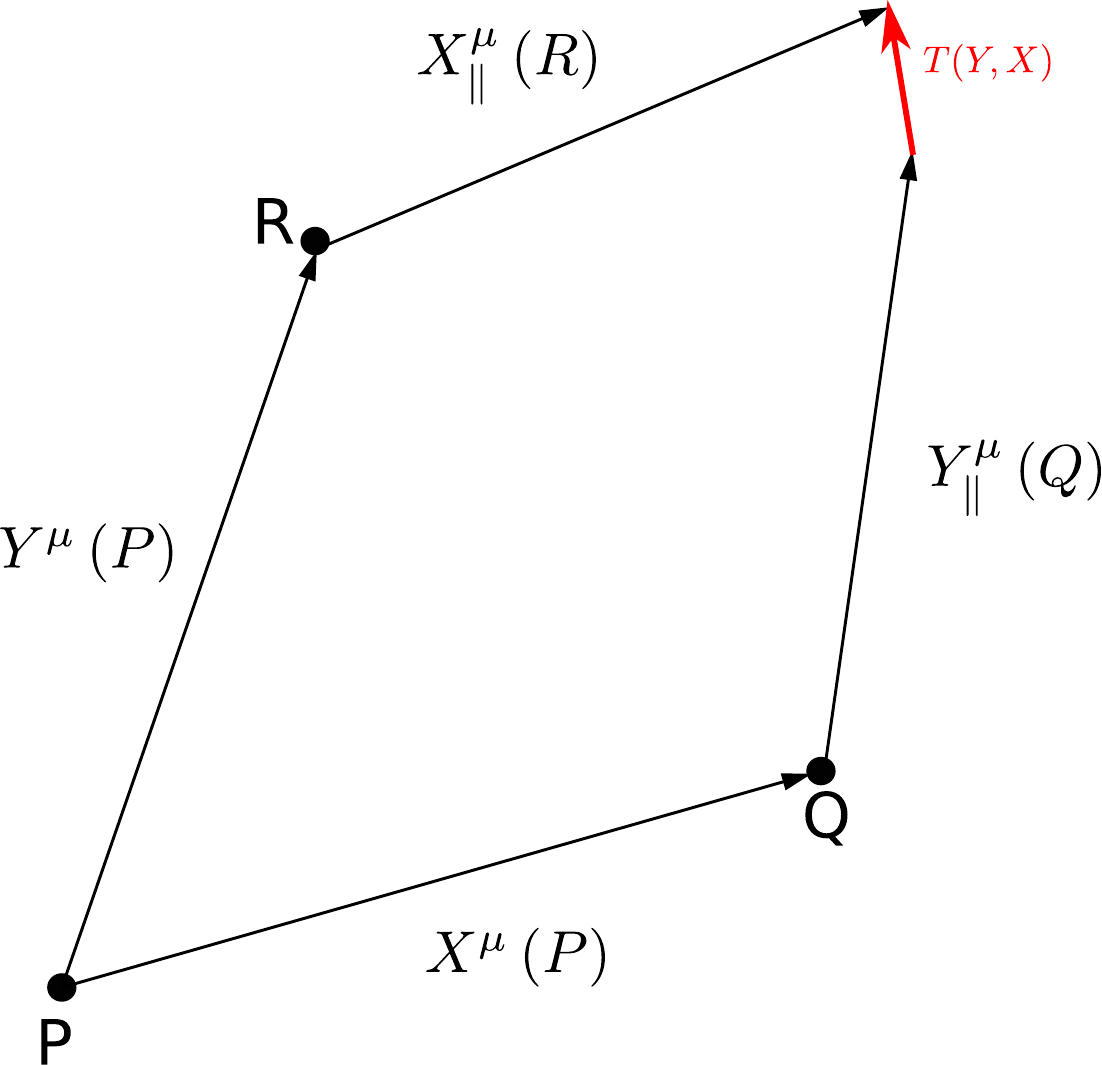}
\caption{One geometric interpretation of torsion in Riemann-Cartan spaces. Consider two vector fields, $X$ and $Y$, at a point $P$. First, parallel-transport $X$ along $Y$ to the infinitesimally close point $R$. Then, again from $P$, parallel-transport $Y$ along $X$ to reach a point $Q$. The failure of the closure of the parallelogram is the geometrical signal of torsion, and its value is the difference between the two resulting vectors $T(X,Y)$. In Riemannian spaces, $V_{n}$, this tensor is assumed to be zero. The picture was inspired by \cite{HehlObukhov2007} but with the notation of \cite{Nakahara}.}\label{FigTorsion}
\end{center}
\end{figure}

A reasonable assumption is that the local distances do not change under parallel transportation. This can be assured if
\begin{equation}\label{metric_compatibility_condition}
\nabla_{\rho}g_{\mu\nu} = 0 \;.
\end{equation}
The condition (\ref{metric_compatibility_condition}) for a linear connection $\Gamma$ is called \emph{metric compatibility}\footnote{As $Q_{\rho\mu\nu}\equiv \nabla_{\rho}g_{\mu\nu}$ is a third-rank tensor, called the \emph{nonmetricity tensor}, we can wonder how far we can go by not assuming this to be zero. In fact, there are some theories where this tensor has a physical interpretation. However, in this work we always take $Q_{\rho\mu\nu}=0$. For more details about the jargon of different spaces, see \cite{HehlObukhov2007}.}. An $n$-dimensional manifold $M$ with a linear connection preserving local distances, i.e. fulfilling condition (\ref{metric_compatibility_condition}), is called a \emph{Riemann-Cartan (RC) space}, denoted by $U_{n}$. Fig. \ref{FigTorsion} gives a geometric interpretation of torsion, with details in the caption. This is the first link between the linear connection and the metric.

GR further assumes that the torsion tensor vanishes, i.e., that the linear connection is symmetric. In such a case, the manifold is called (pseudo-)Riemannian, denoted by $V_{n}$. The unique linear metric-compatible connection without torsion, called the \emph{Levi-Civita connection}, can then be deduced directly from the metric \cite{Nakahara}
\begin{equation}\label{Christoffel_symbols}
\christoffel{\mu}{\nu}{\rho} = \frac{1}{2} \, g^{\mu\sigma} \, ( \partial_{\nu} g_{\rho\sigma} + \partial_{\rho} g_{\nu\sigma} -\partial_{\sigma} g_{\nu\rho} ) \;.
\end{equation}
The quantities (\ref{Christoffel_symbols}) are called \emph{Christoffel symbols}, and the curvature associated with the Levi-Civita connection is the \emph{Riemannian curvature tensor}, denoted by $\tensor{\mathring{R}}{^{\mu}_{\nu\rho\sigma}}$. In this way, the linear connection in a $U_{n}$ space can be written as
\begin{equation*}
\Gamma{^{\mu}}{_{\nu\rho}} = \christoffel{\mu}{\nu}{\rho} + \tensor{K}{^{\mu}_{\nu\rho}} \;,
\end{equation*}
where $\tensor{K}{^{\mu}_{\nu\rho}}\equiv \frac{1}{2}\,(\tensor{T}{^{\mu}_{\nu\rho}} +  \tensor{T}{_{\nu}^{\mu}_{\rho}} + \tensor{T}{_{\rho}^{\mu}_{\nu}})$ is called the \emph{contorsion tensor}\footnote{Some references called it \emph{contortion tensor} \cite{Hehl1976}. However, as we are following closer the terminology of \cite{Nakahara}, we keep the name \emph{contorsion}. As far as we know, there is no consensus yet about the name.}. Notice that,  contrary to the torsion tensor, $\tensor{K}{^{\mu}_{\nu\rho}}$ is not necessarily antisymmetric in the last two indices, unless torsion is totally antisymmetric (see below).

Apart from $V_{n}$ spaces, there are other less standard ``degenerated'' cases of $U_{n}$ spaces. One very known example is the Weitzenb{\"o}ck space \cite{Weitzenbock1923}, where the tensor (\ref{curvature_def}) is assumed to be zero. Still, the Riemannian curvature is not necessarily zero, leading to the \emph{teleparallel gravity} theory \cite{Aldrovandi}. However, we shall not take this direction here.

It is helpful to consider a non-coordinate basis, $e^{a}_{\mu}$, not only because this simplifies the notation, but also because we shall be dealing with spinors. The $e^{a}_{\mu}$, also called \emph{vielbeins}, are defined through the metric as
\begin{equation*}
g_{\mu\nu} = e^{a}_{\mu}\,e^{b}_{\nu} \eta_{ab} \;,
\end{equation*}
where we choose the signature $\eta_{ab}=\mbox{diag}(1,-1, \cdots)$ for the Minkowski metric. The \emph{spin-connection}, $\omega^{ab}_{\mu}=e^{a}_{\lambda}(\delta^{\lambda}_{\nu}\partial_{\mu}+\Gamma^{\lambda}_{\mu\nu})e^{b\nu}$, can be decomposed into torsion-free and contorsion contributions \cite{Nakahara,Z-book}, $\omega_\mu^{ab}=\mathring{\omega}_\mu^{ab}+ \kappa_\mu^{ab}$, where $T^{\lambda}_{\mu\nu}=E_a^{\lambda}\tensor{\kappa}{_{\nu}^{a}_{b}}e^{b}_{\mu}-E_a^{\lambda}\tensor{\kappa}{_{\mu}^{a}_{b}}e^{b}_{\nu}$, being $E_a^{\mu}$ the vielbein inverse, i.e., $e^{a}_{\mu}\,E^{\mu}_{b} = \delta^{a}_{b}$ and $e^{a}_{\mu}\,E^{\nu}_{a} = \delta^{\nu}_{\mu}$.

We are interested in $(2+1)$-dimensional systems so we will focus on $U_{3}$ spaces. A dynamical study of the RC spaces that shows how torsion is related to spin angular momentum can be found in \cite{Hehl1976} and references therein. Here we assume that the geometry is fixed and there are no propagating gravitational degrees of freedom. Nonetheless, we shall keep in mind the natural role of torsion when spinorial fields are at stake \cite{Kibble1961}.  With this, besides possible boundary terms, the graphene-Dirac action with the minimal GR extension in a $U_{3}$ space with a fixed spacetime in a RC geometry is then
\begin{equation}\label{action_torsion}
S = \ii\,\hbar\,v_{F} \int d^{3}x \; |e| \; \overline{\psi}\left(E_{a}^{\mu}\,\gamma^{a}\mathring{\nabla}_{\mu} - \frac{\ii}{4} \gamma^{5} \frac{\epsilon^{\mu\nu\rho}}{|e|} T_{\mu \nu\rho} \right)\psi\;.
\end{equation}
The details of the deduction of each term of (\ref{action_torsion}) are in \cite{ip4}. Let us just mention that, being this the effective action for graphene, here we trade the speed of light, $c$, for the Fermi velocity, $v_F$. Furthermore, as for notation, $|e|=\sqrt{|g|}$, the covariant derivative, $\mathring{\nabla}_{\mu}$, is based on the torsion-free connection, $\mathring{\omega}_\mu^{ab}$, only, and $\gamma^{5} \equiv  i \gamma^{0} \gamma^{1} \gamma^{2} = \left(
                                                           \begin{array}{cc}
                                                             I_{2 \times 2} & 0 \\
                                                             0 & -I_{2 \times 2} \\
                                                           \end{array}
                                                         \right)$
(we used the conventions of \cite{ip3} for $\gamma^{0},\gamma^{1},\gamma^{2}$ giving a $\gamma^{5}$ that {\it commutes} with the other three gamma matrices\footnote{This is due to the reducible, rather than irreducible, representation of the Lorentz group we use.}).

From the action (\ref{action_torsion}), the well-known fact that torsion couples to spinors through its totally antisymmetric component \cite{Shapiro}, can be read-off. Therefore, as remarked above, in this case, the contorsion $\tensor{K}{^{\mu}_{\nu\rho}}$ is also antisymmetric, as can be seen from its definition. This mathematical fact is behind the obstruction pointed out some time ago leading to the conclusion that, in two-dimensional Dirac materials, torsion can have no physical role. On this see, e.g., \cite{deJuan2010,Vozmediano:2010zz,Amorim:2015bga}.

\section{Topological defects and their continuum limit in two-dimensional materials}\label{SecDefects}

This work focuses primarily on graphene-like materials, but many of the following considerations also apply to other two-dimensional crystals. The two most important topological defects in these materials are \emph{disclinations} and \emph{dislocations}\footnote{Another important defect which could have a geometrical meaning, and has possible impacts in using graphene-like materials as analogs, is the \emph{grain boundary} \cite{ip3}.}. These two defects are related to curvature and torsion, respectively, in a continuum-limit description of graphene-like (and many other) materials \cite{Kleinert_book,RuggieroTartaglia2003}.

In a honeycomb lattice, i.e., with a hexagonal structure, the dislocation defect is an $n$-side polygon, with $n=3,\,4,\,5$, or $n=7,\,8\,9\,\ldots$. The defects belonging to the first set ($n \le 5$) carry an intrinsic \textit{positive} curvature, while the defects belonging to the other set ($n \ge 7$) hold intrinsic \textit{negative} curvature. Specific arrangements for such defects are then obeyed when the overall curvature is positive, see, e.g., \cite{GONZALEZ1993771}, and other specific arrangements are obeyed when the overall curvature is negative, see, e.g.,  \cite{LOBACHEVSKYgraphene2016}.

\begin{figure}
\begin{center}
\includegraphics[width=0.65\textwidth,angle=0]{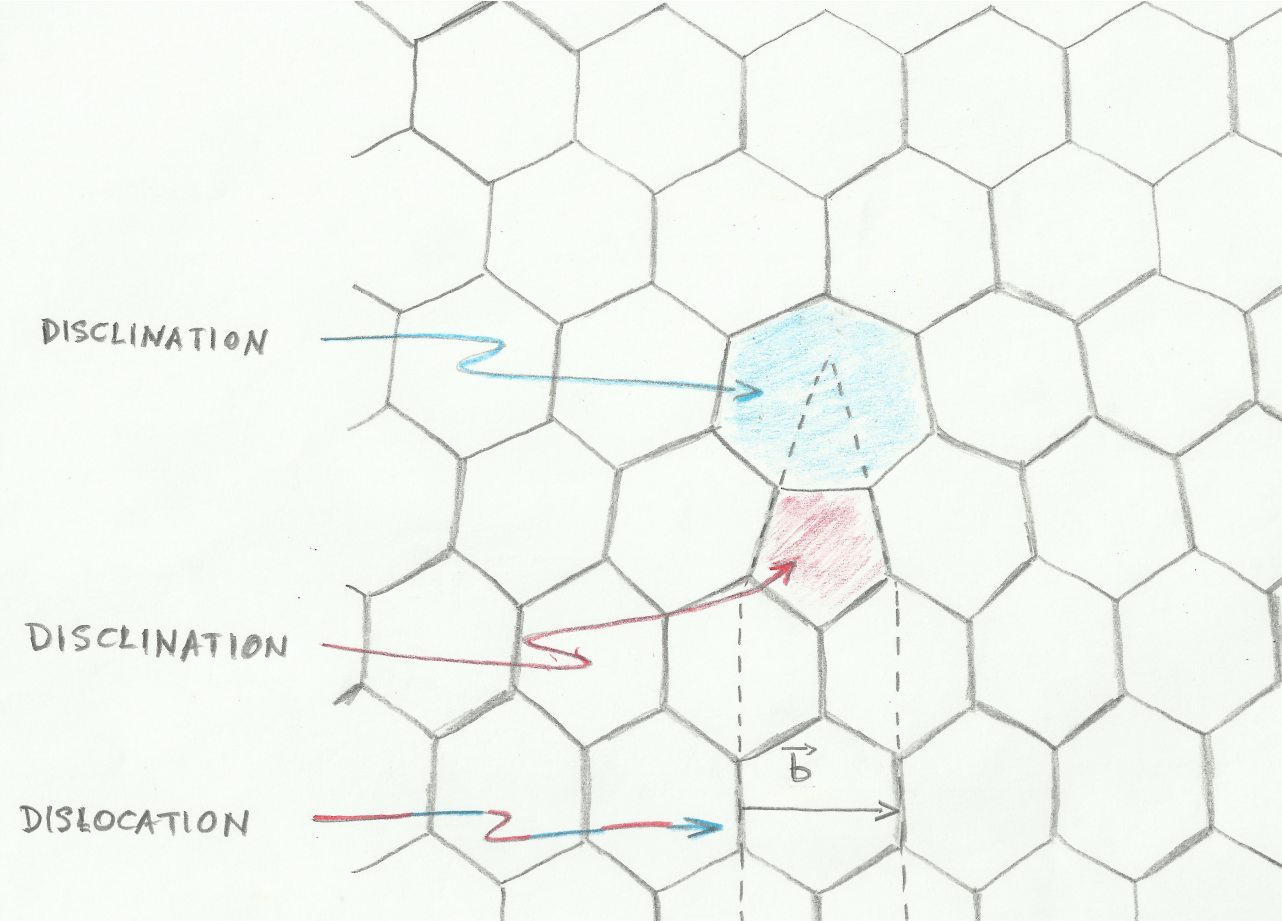}
\caption{Edge dislocation from two disclinations. Two disclinations, a heptagon and a pentagon, by adding up to zero total intrinsic curvature. This makes a dislocation with Burgers vector $\vec{b}$, as can be seen in the lower half-plane. Burgers vector in the continuum limit caries torsion. This figure is taken from \cite{ip3}.}\label{FigEdgeDislocation}
\end{center}
\end{figure}

One way to produce dislocation defects is to have a dipole of disclinations with zero total curvature, as the pentagon-heptagon dipole of Fig.~\ref{FigEdgeDislocation}. This corresponds to introducing a strip in the lower half-plane of such a figure. The \emph{Burgers vector} $\vec{b}$ characterizes this type of defect, and its modulus is the width of the introduced strip. The process of cutting and sewing is called a \emph{Volterra process} \cite{RuggieroTartaglia2003}, and it gives us the intuition of why these defects are topological: one cannot undo them by a continuous transformation.

The torsion tensor in crystals is related to the Burgers vector through the formula \cite{Kleinert_book,Katanaev2005}
\begin{equation}\label{torsion-Burgers}
b^{a}=\int\int_{\Sigma} e^{a}_{\lambda}T^{\lambda}_{\mu \nu}dx^{\mu} \wedge dx^{\nu} \;,
\end{equation}
where $\Sigma$ is a surface containing the dislocation, but otherwise arbitrary, $a = 0,1,2$. However, the apparent simplicity of (\ref{torsion-Burgers}) should not shadow the subtleties that it involves. In fact, given a distribution of Burgers vector, it is not obvious how to assign a specific corresponding torsion tensor \cite{Lazar2003}. See, e.g., \cite{Slager2017,banhart2010structural} and references therein to learn how the dislocation distributions have potential electronic device applications.

From (\ref{torsion-Burgers}), we see that to have a nonzero Burgers vector giving rise to $\epsilon^{\mu\nu\rho}T_{\mu \nu\rho}\neq0$ (which is necessary for a nonzero coupling in (\ref{action_torsion})), we can follow two roads: (i) a \textit{time-directed} screw dislocation (only possible if the crystal has a time direction)
\begin{equation}\label{timecrys}
  b_{t} \propto \int\int  T_{012} dx \wedge dy \,,
\end{equation}
or (ii) an edge dislocation ``felt'' by an integration along a \textit{space-time circuit} (only possible if we can actually go around a loop in time), e.g,
\begin{equation}\label{tloop}
b_{x} \propto \int\int  T_{102} dt \wedge dy \,.
\end{equation}

\section{Torsion through time crystals or through time-loops} \label{SecTimeCrystalsTimeLoops}

The point we want to make here is that scenarios (i) and (ii) are, in fact, not impossible to realize in a laboratory.

Scenario (i) could be explored in the context of the fascinating time crystals introduced by Wilczek \cite{Wilczek2012, WilczekShapere2012}, being nowadays the focus of intense experimental studies \cite{PhysRevLett.109.163001,PhysRevLett.121.185301}. Such lattices, discrete in all dimensions, including time, would be an interesting playground to probe quantum gravity ideas \cite{Loll1998}. In particular, it would be precious to explore defect-based models of classical gravity/geometry, see \cite{Katanaev:1992kh} and \cite{Kleinert_book}. Despite the beauty of scenario (i), we shall focus only on the more manageable but still quite challenging scenario (ii).

Let us assume the Riemann curvature to be zero, $\tensor{\mathring{R}}{^{\mu}_{\nu\rho\sigma}}=0$, but torsion (or contorsion $\tensor{K}{^{\mu}_{\nu\rho}}\neq0$) to be nonzero, and let us choose a frame where $\mathring{\omega}_\mu^{ab}=0$. The action (\ref{action_torsion}) is then
\begin{equation}\label{action_pure_torsion}
S = \ii\,\hbar\,v_F \int d^{3}x |e| \,  \left(\overline{\psi}\gamma^{\mu}\partial_{\mu}\psi - \frac{\ii}{4} \overline{\psi}_{+} \phi \psi_{+} + \frac{\ii}{4} \overline{\psi}_{-} \phi \psi_{-} \right) \;,
\end{equation}
where $\psi = (\psi_+,\psi_-)$ and $\phi \equiv \epsilon^{\mu \nu \rho} T_{\mu\nu\rho} / |e|$ is what we call \emph{torsion field}. Even in the presence of torsion, the two irreducible spinors, $\psi_{+}$ and $\psi_{-}$, are decoupled (however, they couple to $\phi$ with opposite signs).

To overcome the three-dimensional geometric obstruction through the ``time-loop'' in the $(y,t)$-plane of scenario (ii), see (\ref{tloop}), our proposal is to make use of the particle-antiparticle description of the dynamics encoded in the action (\ref{action_pure_torsion}). The picture goes as follows.

First, one realizes that the regime of the materials we are describing is the ``half-filling'' \cite{pacoreview2009}, for which the energy states of the valence band ($E < 0$) have the vacancies completely filled (being the analog of the Dirac sea), while the vacancies of the conduction band ($E > 0$) stay empty.

\begin{figure}
\begin{center}
\includegraphics[width=0.4\textwidth,angle=90]{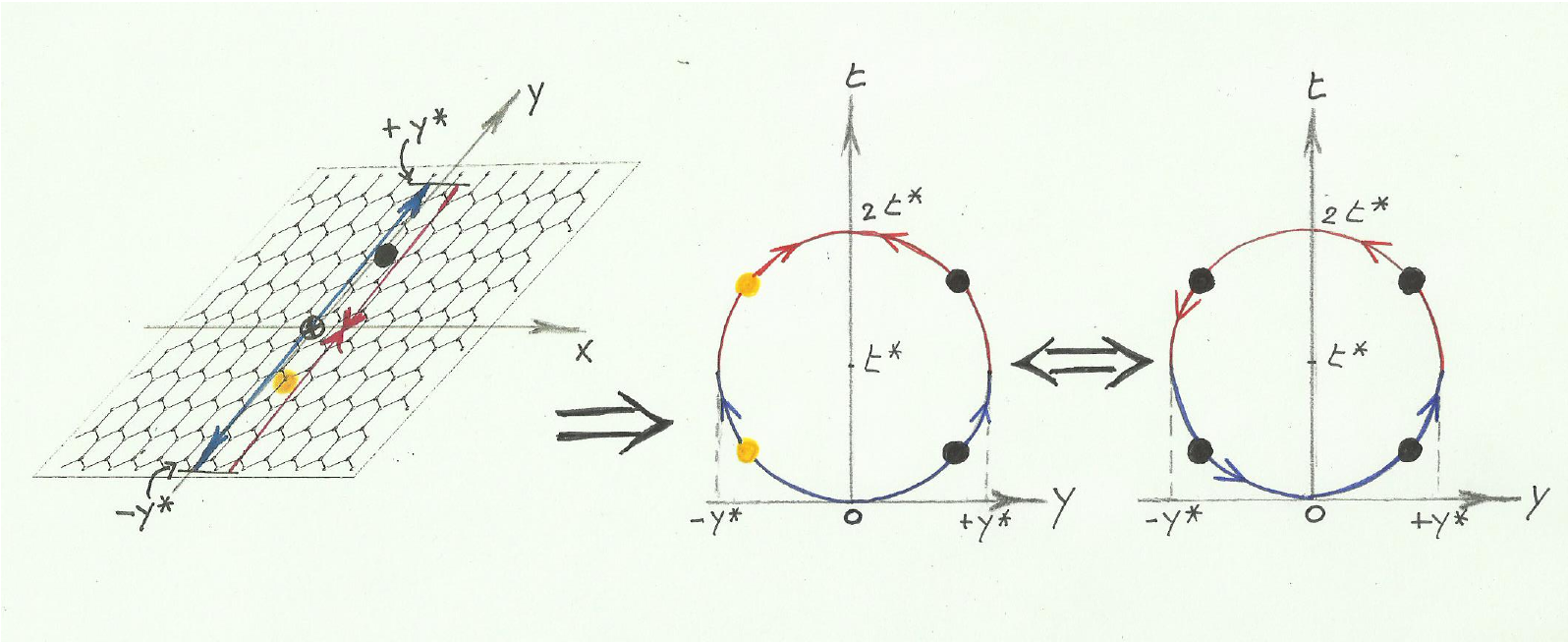}
\end{center}
\caption{Idealized \emph{time-loop}. At $t=0$, the hole (yellow) and the particle (black) start their journey from $y=0$, in opposite directions. Evolving forward in time, at $t=t^*>0$, the hole reaches $-y^*$, while
the particle reaches $+y^*$,  (blue portion of the circuit). Then they come back to the original position, $y=0$, at $t=2t^*$ (red portion of the circuit). This can be repeated indefinitely. On the far right, the
equivalent \textit{time-loop}, where the hole moving forward in time is replaced by a particle moving backward in time. Figure taken from \cite{ip4}.}%
\label{Fig3TIMELOOP}%
\end{figure}

Think now of exciting a pair particle-hole out of this vacuum and making them oscillate, say, along the $y$-axis, as described in Fig.~\ref{Fig3TIMELOOP}. This amounts to a circuit of the particle-antiparticle pair in the $(y,t)$-plane. What is left to do is to fully exploit the emergent relativistic-like structure of the model and see the portion of the circuit described by the \textit{antiparticle} moving \textit{forwards} in time, as corresponding to the same \textit{particle} moving \textit{backwards} in time. This realizes what we may call a \textit{time-loop}.

The pictures in Fig.~\ref{Fig3TIMELOOP} refer to a defect-free honeycomb graphene-like sheet. The presence of a dislocation, with Burgers vector $\vec{b}$ directed along $x$, would result in a failure to close the loop proportional to $\vec{b}$ \cite{ip4}.

Therefore, going from a first to a second quantization approach, paying due attention to the subtleties involved in describing dislocation distributions by a suitable torsion tensor, the Dirac field theory emerging here can indeed include a nonzero coupling with torsion. This accounts for a field theoretical description of the effects of dislocations only when the third dimension is taken to be time, overcoming the geometrical obstruction discussed earlier \cite{deJuan2010,Vozmediano:2010zz,Amorim:2015bga}.

\section{Linear and non-linear response regimes}

The above is a nice idea, but the real challenge is to bring this idealized picture close to experiments. We present below the first steps in that direction; see \cite{ip4} for more details.

The simplest settings we can envisage to realize the picture above-presented need: i) an \emph{external electromagnetic field} to excite the particle-hole pair necessary for the time-loop, and ii) that a suitable disclination/torsion provides the non-closure of the loop in the appropriate direction, something we shall refer to as \emph{holonomy}. In other words, we are looking for \textit{the measurable effects of a disclination/torsion-induced holonomy in a time-loop}. It is only a \textit{suitable combination} of those interactions that can produce the effect we are looking for. Notice this is more related to the \emph{Cartan circuit} for torsion and curvature visualization \cite{HehlObukhov2007} than the non-closing parallelogram torsion interpretation given in Fig. \ref{FigTorsion}.

Therefore, the action governing the relevant microscopic dynamics is
\begin{eqnarray}
S  & = & \ii  \int d^{3}x \,|e|\, \left(\overline{\Psi} \gamma^{\mu} (\partial_{\mu} - \ii g_{\mbox{em}}\, A_\mu) \Psi - \ii\, g_{\mbox{tor}}\, \overline{\psi}_{+} \phi \psi_{+} + \ii\,g_{\mbox{tor}}\,\overline{\psi}_{-}\,\phi\,\psi_{-} \right)  \label{action torsion external A} \\
& \rightarrow & \ii  \int d^{3}x \, \left(\overline{\psi} \gamma^{\mu} \partial_{\mu} \psi  - \ii\,g_{\mbox{em}}\,\,\hat{j}^{\mu}_{\mbox{em}}\,A_\mu - \ii\,g_{\mbox{tor}}\,\hat{j}_{\mbox{tor}}\,\phi \right) \equiv S_0 [\overline{\psi}, \psi] + S_I [A, \phi] \;. \label{action samples}
\end{eqnarray}
where we have set constants to one, $g_{\mbox{em}}$ and $g_{\mbox{tor}}$ are the electromagnetic and torsion coupling constants, respectively. In the last line, to avoid unnecessary complications, we considered only one Dirac point, say $\psi \equiv \psi_+$, and the metric is taken to be flat, $|e|=1$. Hence the indices are the flat ones, $\mu, \nu, ... \to a,b,...$, but to ease the notation, we shall use Greek letters, anyway. Finally, $\hat{j}^{\mu}_{\mbox{em}} \equiv  \overline{\psi} \gamma^{\mu} \psi$, while $\hat{j}_{\mbox{tor}} \equiv \overline{\psi} \psi$.

The electromagnetic field is \textit{external}, hence a four-vector\footnote{A different, if not more natural $(2+1)-$dimensional setting would be to obtain $A_\mu$ by suitably straining the material, see, e.g., \cite{Vozmediano:2010zz,Amorim:2015bga}, and \cite{ipfirst}. In that case, a typical setting is $A_t \equiv 0$, $A_x \sim u_{xx} - u_{yy}$, $A_x \sim 2 u_{xy}$, where $u_{i j}$ is the strain tensor.} $A_\mu \equiv (V, A_x, A_y,A_z)$. Nonetheless, the dynamics it induces on the electrons living on the membrane is two-dimensional. Therefore, the effective vector potential may be taken to be $A_\mu \equiv (V, A_x, A_y)$, see, e.g., \cite{PhysRevLett.121.207401, natureLaserGraphene}. Alternatively, the so-called \emph{reduced QED} approach can be taken \cite{Marino:1992xi,Gorbar:2001qt}. In such an approach, the gauge field propagates in a three-dimensional space and one direction is integrated out to obtain an effective interaction with the electrons constrained to move in a two-dimensional plane.\footnote{This approach could shed some light on the appearance of a photon Chern-Simons term. On this, see \cite{Dudal:2018mms,Dudal:2018pta}.}

As mentioned above, we do not consider the dynamics of defects here. Hence the torsion field $\phi$ as well enters into the action as an external field. A different view, when $\phi$ is constant, is to include it into the unperturbed action, where it plays the role of a mass $S_0 \to S_m$, see, e.g., \cite{DauriaZanelli2019}, where $S_m = i \int d^{3}x  \; \overline{\psi}(\slashed{\partial} - m(\phi)) \psi$.

We are in the situation described by the microscopic perturbation
\begin{equation}\label{SIfx}
S_I [F_i] = \int d^{3}x \, \hat{X}_i (\vec{x},t) F_i (\vec{x},t) \;,
\end{equation}
with the system responding through $\hat{X}_i (\vec{x},t)$ to the external probes $F_i (\vec{x},t)$. The general goal is then to find
\begin{equation}\label{X[F]}
\hat{X}_i [F_i] \;,
\end{equation}
to the extent of predicting a measurable effect of the combined action of the two perturbations $F_i (\vec{x},t)$: $F^{\mbox{em}}_{1} (\vec{x},t) \propto A_\mu (\vec{x},t)$ that induces the response $\hat{j}^\mu_{\mbox{em}}$, and $F^{\mbox{tor}}_{2} (\vec{x},t) \propto \phi (\vec{x},t)$ that induces the response  $\hat{j}_{\mbox{tor}}$:
\begin{equation}\label{SIaphi}
S_I [A, \phi] = \int d^{3}x \, \left( \hat{j}^\mu_{\mbox{em}} A_\mu + \hat{j}_{\mbox{tor}} \phi \right) \,,
\end{equation}
where we have included the couplings, $g_{\mbox{em}}$ and $g_{\mbox{tor}}$, in the respective currents.

In fact, in our model, described microscopically by the action \eqref{action samples}, we can indeed produce a prediction based on the charge conjugation invariance of that emergent relativistic theory. Such prediction is that
\begin{equation}\label{firstFurry}
    \chi^{\mbox{torem}}_{\mu} (x,x') \sim \langle \hat{j}^{\mbox{em}}_{\mu} (x) \hat{j}^{\mbox{tor}} (x') \rangle \equiv 0 \;.
\end{equation}
This is nothing more than an instance of the Furry's theorem of quantum field theory \cite{Peskin}, that in QED reads
\begin{equation}\label{FurryEMgeneral}
    \chi^{\mbox{em}}_{\mu_{1} ... \mu_{2n+1}} (x_{1}, ..., x_{2n+1}) \sim \langle \hat{j}^{\mbox{em}}_{\mu_{1}} (x_{1}) \cdots \hat{j}^{\mbox{em}}_{\mu_{2n+1}} (x_{2n+1}) \rangle = 0 \;,
\end{equation}
and for us implies
\begin{equation}\label{Furry_ours}
    \chi^{\mbox{torem}}_{\mu_{1} ... \mu_{2n+1}} (x_{1}, ..., x_{2n+1}, y_{1}, ..., y_{m}) \sim \langle \hat{j}^{\mbox{em}}_{\mu_{1}} (x_{1}) \cdots \hat{j}^{\mbox{em}}_{\mu_{2n+1}} (x_{2n+1})
    \hat{j}^{\mbox{tor}} (y_1) \cdots \hat{j}^{\mbox{tor}} (y_{m})\rangle = 0 \;.
\end{equation}

This result tells us that the effects we are looking for can only be seen if we move to the nonlinear response regime. We can resort to a well-developed technique, the high-order harmonic generation (HHG), which can characterize structural changes both in atoms and molecules and, more recently, bulk materials (for a recent review, see, e.g.~\cite{StanislavRevModPhys}). Therefore, in our scheme, the intra-band harmonics, governed by the intra-band (electron-hole) current, will be strongly modified, depending on whether dislocations are there or not.

\section{Conclusions}

When time is duly included in the emergent relativistic-like picture of Dirac materials, the geometric obstruction to describe the effects of dislocations in terms of a suitable coupling with torsion within the $(2+1)$-dimensional field theoretical description of the $\pi$-electrons dynamics can be overcome. However, surely problems remain to be addressed, like a unique assignment of torsion to a given distribution of Burgers vectors. Nonetheless, when this is possible, our suggestion here opens the doors to using these materials as analogs of many important theoretical scenarios where torsion plays a role.

We then show the first steps toward testing the realization of these scenarios. We have focussed on one of the \textit{Gedankenexperiments} of \cite{ip4} on \textit{time-loop}. This could spot the presence of edge dislocations, routinely produced in Dirac materials. The effect must be based on the interplay between an external electromagnetic field (necessary to excite the pair particle-hole that realizes the time-loops) and a suitable distribution of dislocations described as torsion (that will be responsible for a measurable holonomy in the time-loop).

We also can prove, on very general grounds, that a nonlinear response regime is necessary. In particular, in an HHG technique, the structure of such a response would include manifestation of the torsion-induced holonomy of the time-loop through specific suppression patterns and generation of higher harmonics. This sounds promising, for an experimental finding, as the laser-graphene interaction, controlling electron dynamics on an unprecedented precision scale, is the focus of intense studies, both theoretical and experimental, see, e.g., \cite{PhysRevLett.121.207401, natureLaserGraphene}.

\section*{Acknowledgements}

We sincerely thank Thomas Elze and the other organizers for putting together an amazing conference. We are indebted to Marcelo Ciappina and Adamantia Zampeli for their collaboration on these exciting matters and Jorge Zanelli for many fruitful inspirational discussions. We gladly acknowledge support from Charles University Research Center (UNCE/SCI/013). P.~P. is also supported by Fondo Nacional de Desarrollo Cient\'{i}fico y Tecnol\'{o}gico--Chile (Fondecyt Grant No.~3200725).

\printbibliography

\end{document}